\documentstyle[aps,preprint,tighten,epsf]{revtex}
\begin{document}
\title{Using projections and correlations to \\
       approximate probability distributions}
\author{Dean Karlen\footnote{E-mail: karlen@physics.carleton.ca}}
\address{Ottawa-Carleton Institute for Physics\\
         Department of Physics,
         Carleton University\\
         Ottawa, Canada K1S 5B6}
\date{May 7, 1998}
\maketitle
\begin{abstract}
A method to approximate continuous
multi-dimensional probability density functions
(PDFs) using their projections and correlations is described.
The method is particularly useful for event classification when 
estimates of systematic uncertainties are required and
for the application of an unbinned maximum likelihood
analysis when an analytic model is not available.
A simple goodness of fit test of the approximation can be used,
and simulated event samples that follow the approximate PDFs
can be efficiently generated.
The source code for a FORTRAN-77 implementation of this method is
available.
\end{abstract}
\pacs{ }


\section{Introduction}

Visualization of multi-dimensional distributions is often performed by
examining
single variable distributions (that is, one-dimensional projections) and
linear correlation coefficients amongst the variables.
This can be adequate when the sample size is small, 
the distribution consists of essentially
uncorrelated variables, or when the correlations between the variables is
approximately linear.
This paper describes a method to approximate multi-dimensional
distributions in this manner and its applications in data analysis.

The method described in this paper, the
Projection and Correlation Approximation (PCA),
is particularly useful in analyses which make use of either simulated or
control event samples.
In particle physics,
for example, such samples are used to develop algorithms
that efficiently select events of one type while
preferentially rejecting events of other types.
The algorithm can be as simple as a set of criteria on quantities
directly measured in the experiment or as complex as an application of
an artificial neural network \cite{REF:ann}
on a large number of observables.
The more complex algorithm may result in higher efficiency and
purity, but the determination of systematic errors
can be difficult to estimate.
The PCA method can be used to define a
sophisticated selection algorithm with good efficiency and purity,  
in a way that systematic uncertainties can be reliably estimated.

Another application of the PCA method
is in parameter estimation from a data set using a
maximum likelihood technique.
If the information available is in the form of
simulated event samples, it 
can be difficult to apply an unbinned maximum likelihood method, 
because it requires
a functional representation of
the multidimensional probability density function (PDF).
The PCA method can be used to approximate the
PDFs required for the maximum likelihood method.
A simple goodness of fit test is available to determine if the
approximation is valid.

To verify the statistical uncertainty of an analysis, it can
be useful to create a large ensemble of simulated samples, each
sample equivalent in size to the data set being analyzed.
In cases where this is not practical because of limited computing resources,
the approximation developed in the PCA method can be used,
as it is in a form that leads to an efficient method for event generation.

In the following sections, the projection and correlation approximation
will be described along with its applications.
An example data analysis using the PCA method is shown.

\section{Projection and correlation approximation}\label{SECT:pca}

Consider an arbitrary probability density function ${\cal P}(\bf x)$ of
$n$ variables, $x_i$.
The basis for the approximation of this PDF
using the PCA approach is the
$n$-dimensional Gaussian distribution, centered at the origin,
which is described by
an $n\times n$ covariance matrix, $V$, by
\begin{equation}
  G({\bf y}) = (2\pi)^{-n/2}\,\vert V\vert^{-1/2} 
      \exp\left({-{{\textstyle{1\over2}}}\, {\bf y}^T\, V^{-1}\, {\bf y}}
          \right)
  \label{EQ:ngaus}
\end{equation}
where $\vert V \vert$ is the determinant of $V$.
The variables ${\bf x}$ are not, in general, Gaussian distributed
so this formula would be a poor approximation of the PDF, if used directly.
Instead, the PCA method uses parameter transformations, $y_i(x_i)$,
such that the individual distributions for $y_i$ are Gaussian 
and, as a result, the
$n$-dimensional distribution for ${\bf y}$ may be well approximated
by Eq.\ (\ref{EQ:ngaus}).

The monotonic function $y(x)$ that transforms a variable $x$, having
a distribution function $p(x)$, to the variable $y$, which follows
a Gaussian distribution of mean~0 and variance~1, is
\begin{equation}
y(x)=\sqrt{2}\,{\rm erf}^{-1}\left(2F(x)-1 \right)
\label{EQ:transform}
\end{equation}
where erf$^{-1}$ is the inverse error function and
$F(x)$ is the cumulative distribution of $x$,
\begin{equation}
F(x) = {\int_{x_{\rm min}}^x p(x')\,dx' \over
        \int_{x_{\rm min}}^{x_{\rm max}} p(x')\,dx'} \ \ .
\label{EQ:cumul}
\end{equation}

The resulting $n$-dimensional
distribution for {\bf y} will not, in general, be an $n$-dimensional
Gaussian distribution.
It is only guaranteed that the projections of this distribution onto each
$y_i$ axis is Gaussian.
In the PCA approximation, however,
the probability density function of {\bf y}
is assumed to be Gaussian.
Although not exact, this can represent a good approximation
of a multi-dimensional distribution in which the correlation of the
variables is relatively simple.

Written in terms of the projections, $p_i(x_i)$,
the approximation of ${\cal P}({\bf x})$ using the PCA method is,
\begin{equation}
P({\bf x})= \vert V \vert^{-1/2}
          \exp\left(-{{\textstyle{1\over2}}}\,{\bf y}^T\,(V^{-1}-I)\,{\bf y}
              \right)
              \prod_{i=1}^n p_i(x_i)
\label{EQ:pca}
\end{equation}
where $V$ is the covariance matrix for {\bf y} and
$I$ is the identity matrix.
To approximate the projections, $p_i(x_i)$, needed in Eqs.\
(\ref{EQ:cumul}) and (\ref{EQ:pca}),
binned frequency distributions (histograms)
of $x_i$ can be used.

The projection and correlation approximation is exact for distributions
with uncorrelated variables, in which case $V=I$.
It is also exact for a Gaussian distribution modified by
monotonic one-dimensional variable transformations for any 
number of variables; or equivalently,
multiplication by a non-negative separable function.

A large variety of distributions can be well approximated by
the PCA method.
However, there are distributions for which this will not be true.
For the PCA method to yield a good approximation
in two-dimensions, the correlation between
the two variables must be the same sign for all regions.
If the space can be split into regions,
inside of which the correlation has everywhere the same sign, 
then the PCA method can be used on each region separately.
To determine if a distribution is well approximated by the PCA method,
a goodness of fit test can be applied, as described in the next section.

The generation of simulated event samples that follow the PCA PDF is
straightforward and efficient.
Events are generated in $y$ space, according to
Eq.\ (\ref{EQ:ngaus}),
and then are transformed to the $x$ space.
The procedure involves no rejection of trial events, and is therefore
fully efficient.

\section{Goodness of fit test}\label{SECT:goodness}

Some applications of the PCA method
do not require that the
PDFs be particularly well approximated.
For example, to
estimate the purity and efficiency of event classification, it is only
necessary that the simulated or control
samples are good representations of the data.
Other applications, such as its use in maximum likelihood analyses,
require the PDF to be a good approximation, in order that the estimators
are unbiased and that the estimated statistical uncertainties are valid.
Therefore it may be important to check that the approximate PDF
derived with the PCA method is adequate for a given problem.

In general, when approximating a multidimensional distribution from a
sample of events, 
it can be difficult to derive a goodness of fit
statistic, like a $\chi^2$ statistic.
This is because the required multidimensional binning can
reduce the average number of events per bin to a very small number,
much less than 1.

When the PCA method is used, however, it is easy to form a statistic
to test if a sample of events follows the PDF, without 
slicing the variable space into thousands of bins.
The PCA method already ensures that the projections of the approximate
PDF will match that of the event sample. 
A statistic that is sensitive to the correlation amongst the variables
is most easily defined in
the space of transformed variables, $y$, where the approximate PDF
is an $n$-dimensional Gaussian.
For each event the value $X^2$ is calculated,
\begin{equation}
X^2 = {\bf y}^T\,V^{-1}\,{\bf y} \ \ ,
\end{equation}
and if the events follow the PDF, the $X^2$ values 
will follow a $\chi^2$ distribution
with $n$ degrees of freedom, where $n$ is the
dimension of the Gaussian.
A probability weight, $w$, can therefore be formed,
\begin{equation}
w(X^2) = \int_{X^2}^\infty\,\chi^2(t,n)\,dt \ \ ,
\end{equation}
which will be uniformly
distributed between 0 and 1, if the events follow the PDF.
The procedure can be thought of in terms of dividing the $n$-dimensional
$y$ space into layers centered about the origin (and whose
boundaries are at constant probability in $y$~space)
and checking that the right number of events appears in each layer.
The goodness of fit test for the PCA distribution
is therefore reduced to a test that the 
$w$ distribution is uniform.

When the goodness of fit test shows that the event sample
is not well described by the projection and correlation
approximation, further steps may be
necessary before the PCA method can be applied to an analysis.
To identify correlations which are poorly described, the goodness of
fit test can be repeated for each pair of variables.
If the test fails for a pair of variables, it may be possible to
improve the approximation by modifying the choice of variables
used in the analysis, or by treating different regions of variable
space by separate approximations.

\section{Event classification}\label{SECT:class}

Given two categories of events that follow the PDFs
${\cal P}_1({\bf x})$ and
${\cal P}_2({\bf x})$, the optimal event classification scheme
to define a sample enriched in type 1 events,
selects events having the largest values for the ratio of probabilities,
$R={\cal P}_1({\bf x})/{\cal P}_2({\bf x})$.
Using simulated or control samples, the
PCA method can be used to define the approximate PDFs
$P_1({\bf x})$ and $P_2({\bf x})$, and in order to define a quantity 
limited to the range $[0,1]$, it is useful to define 
a likelihood ratio
\begin{equation}
{\cal L}={ P_1({\bf x}) \over
           P_1({\bf x}) + P_2({\bf x}) } \ .
\label{EQ:probrat}
\end{equation}
With only two categories of events, it is irrelevant if
the PDFs $P_1$ and $P_2$ are renormalized
to their relative abundances in the data set.
The generalization to more than two
categories of events requires that the
PDFs $P_i$ be renormalized to their abundances.
In either case, each event is classified on the basis of the
whether or not the value of ${\cal L}$ for that event
is larger than some critical value.

Systematic errors in the estimated purity and efficiency of
event classification can result if the simulated (or control) samples
do not follow the true PDFs.
To estimate the systematic uncertainties of the selection,
the projections and covariance matrices used to define the PCA 
PDFs can be varied over suitable ranges.

\section{Example application}

In this section the PCA method and its applications are
demonstrated with simple analyses of simulated event samples.
Two samples, one labeled signal and the other background,
are generated with, $x_1\in(0,10)$ and $x_2\in(0,1)$, according
to the distributions,
\begin{eqnarray}
d_s(x_1,x_2)&=&{\displaystyle{{(x_1-a_1)^2+a_2 \over
       (a_3(x_1-a_4(1+a_5x_2))^4+a_6)((x_2-a_7)^4+a_8)}}} \cr
 & & \cr
d_b(x_1,x_2)&=&{\displaystyle{{1\over(b_1(x_1+x_2)^2+b_2x_2^3+b_3)}}}
\label{EQ:mcdist}
\end{eqnarray}
where the vectors of constants are given by
{\bf a}$=(7,2,6,4,0.8,40,0.6,2)$ and
{\bf b}$=(0.1,3,0.1)$.
These samples of 4000 events each
correspond to simulated or control samples
used in the analysis of a data set.
In what follows it is assumed that the analytic forms of the
parent distributions, Eq.\ (\ref{EQ:mcdist}), are unknown.

\begin{figure}
 \begin{center}
   \mbox{
    \epsfysize=16.0cm
    \epsffile[86 424 446 783]{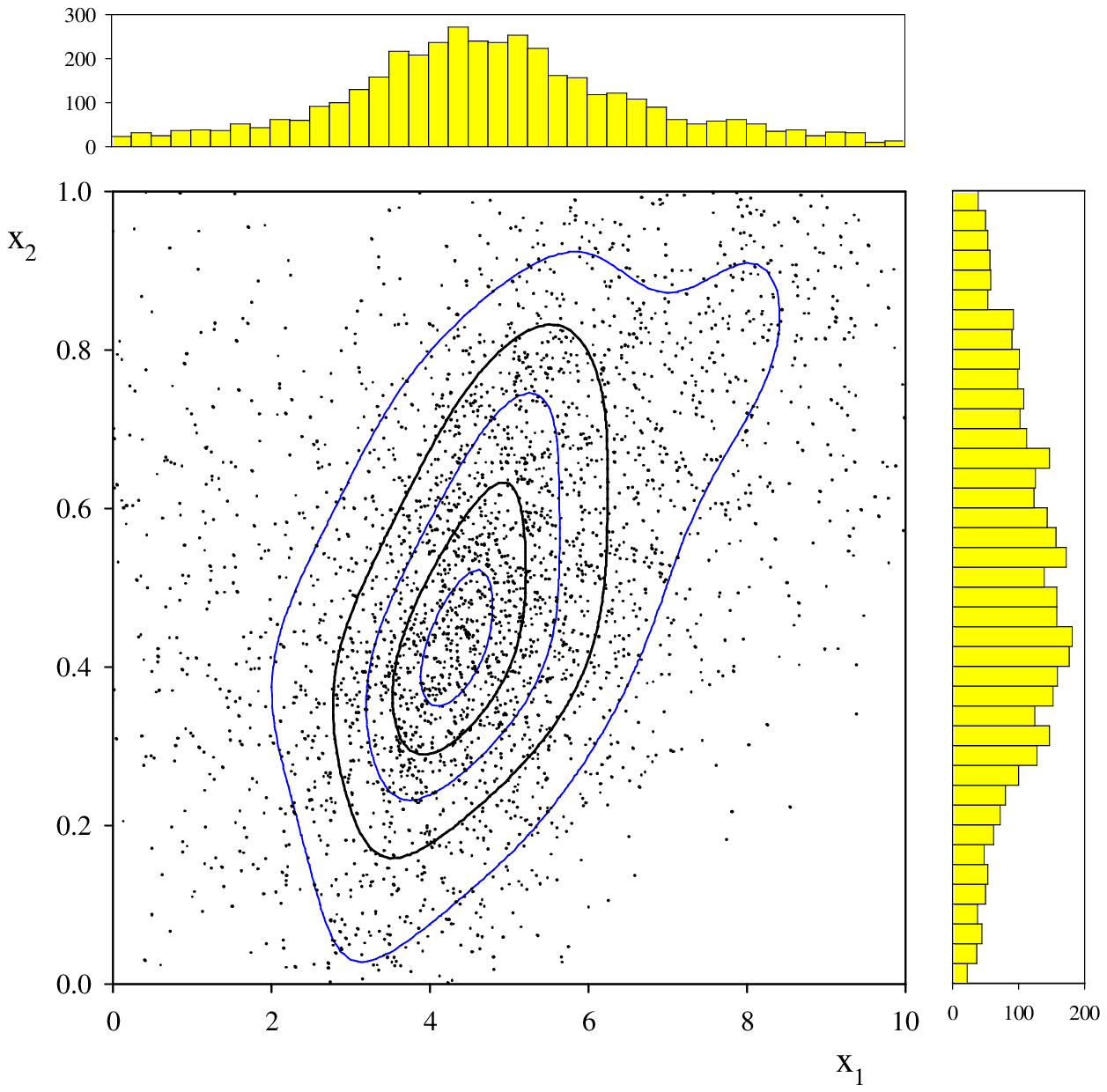}
   }
 \end{center}
\caption[foo]{\label{FIG:sigcontrol}
The points represent the
sample of 4000 events generated according to the function $d_s$
in Eq. (\ref{EQ:mcdist}), which are used as a control sample for
the signal distribution.
Contours of $d_s$ are shown to aid the eye.
The two projections of the distribution are used by the PCA method to
approximate the signal PDF.
}
\end{figure}

\begin{figure}
 \begin{center}
   \mbox{
    \epsfysize=16.0cm
    \epsffile[86 424 446 783]{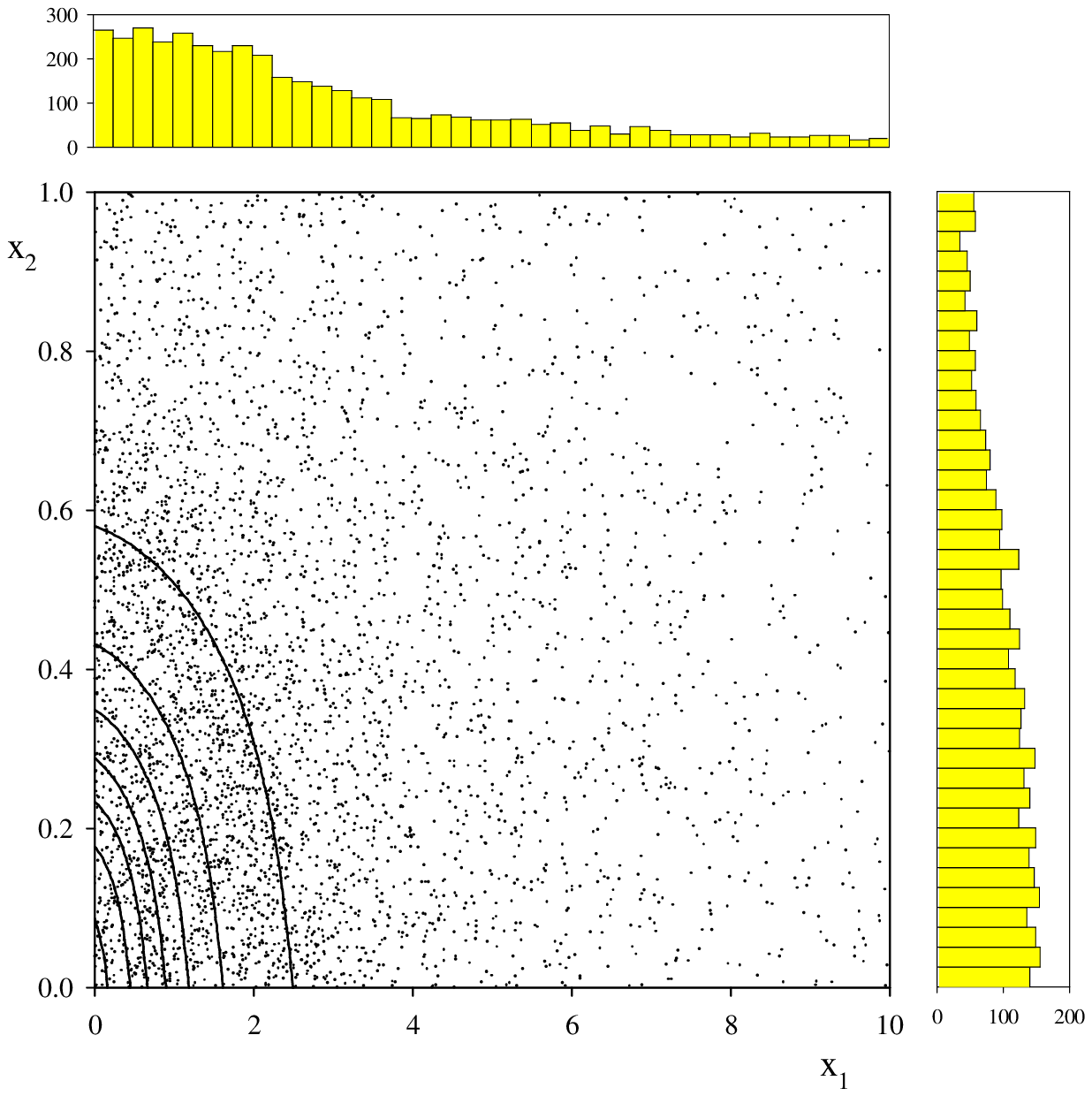}
   }
 \end{center}
\caption[foo]{\label{FIG:backcontrol}
The points represent the
sample of 4000 events generated according to the function $d_b$
in Eq. (\ref{EQ:mcdist}), which are used as a control sample for
the background distribution.
Contours of $d_b$ are shown to aid the eye.
The two projections of the distribution are used by the PCA method to
approximate the background PDF.
}
\end{figure}

\begin{figure}
 \begin{center}
   \mbox{
    \epsfysize=16.0cm
    \epsffile[86 424 446 783]{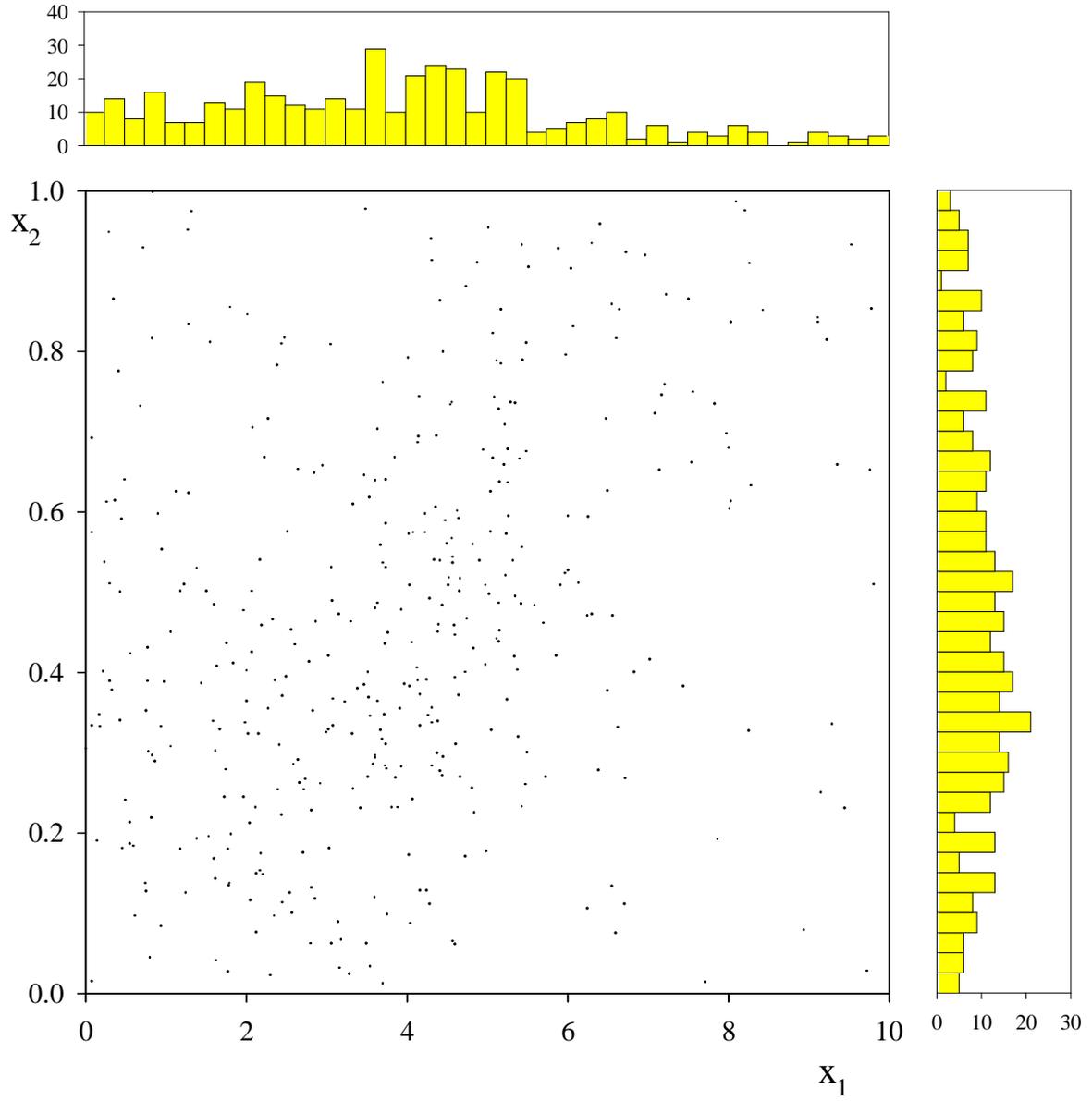}
   }
 \end{center}
\caption[foo]{\label{FIG:data}
The points represent the data sample of 400 events 
consisting of 240 events generated according to the function $d_s$
and 160 generated according to $d_b$
in Eq. (\ref{EQ:mcdist}).
}
\end{figure}

The signal and background control samples are shown in 
Fig.\ \ref{FIG:sigcontrol} and Fig.\ \ref{FIG:backcontrol} respectively. 
A third sample, considered to be data and shown in Fig.\ \ref{FIG:data}, 
is formed
by mixing a further 240 events generated according to $d_s$ and
160 events generated according to $d_b$.

\begin{figure}
 \begin{center}
   \mbox{
    \epsfysize=16.0cm
    \epsffile[86 424 446 783]{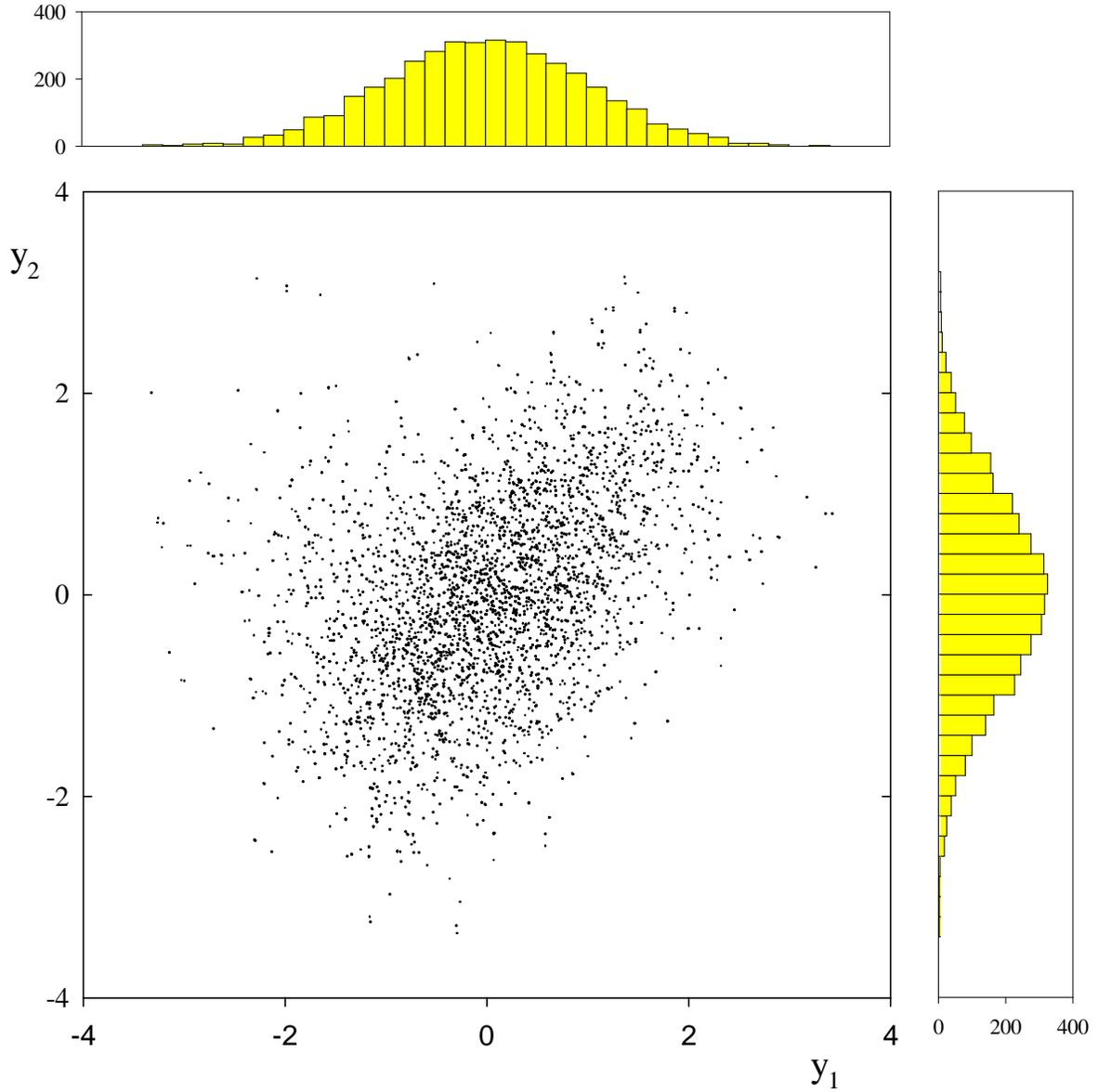}
   }
 \end{center}
\caption[foo]{\label{FIG:sigtrans}
The points show the distribution of the 4000 signal events after
being transformed according to Eq.\ (\ref{EQ:transform}).
The projections are now Gaussian distributions, centered at 0 with
width 1, and the overall distribution appears to follow a 2-dimensional
Gaussian.
The correlation coefficient is 0.40.
}
\end{figure}

\begin{figure}
 \begin{center}
   \mbox{
    \epsfysize=10.0cm
    \epsffile[80 240 490 555]{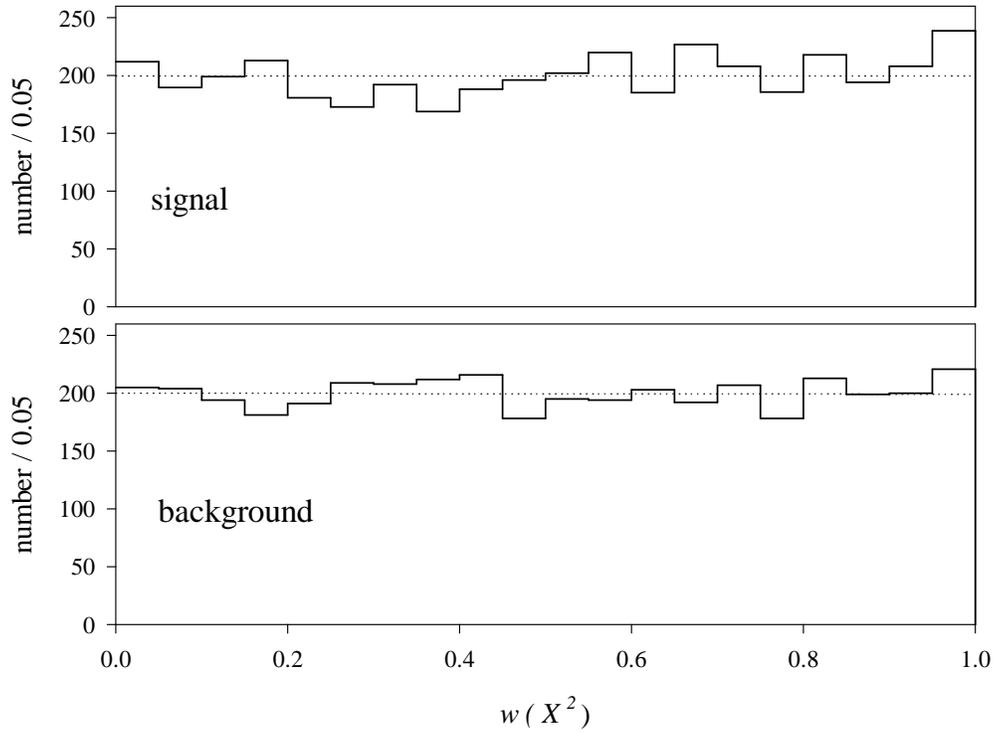}
   }
 \end{center}
\caption[foo]{\label{FIG:wdist}
The upper and lower histograms show the
results of the goodness of fit test applied to the signal and
background control samples.
The $\chi^2$ values are 31 and 14 for 19 degrees of freedom,
respectively.
}
\end{figure}

The transformation given in Eq.\ (\ref{EQ:transform}) is applied to
the signal control sample, which results in the distribution 
shown in Fig.\ \ref{FIG:sigtrans}.
To define the transformation, the projections shown in 
Fig.\ \ref{FIG:sigcontrol} are used, 40 bins for each
dimension.
The projections of the transformed distribution are Gaussian, 
and the correlation
coefficient is found to be 0.40.
The goodness of fit test, described in section \ref{SECT:goodness},
checks the assumption that the transformed
distribution is a 2-dimensional
Gaussian. 
The resulting
$w(X^2)$ distribution from this test is relatively uniform, as shown in
Fig.\ \ref{FIG:wdist}.

\begin{figure}
 \begin{center}
   \mbox{
    \epsfysize=16.0cm
    \epsffile[86 424 446 783]{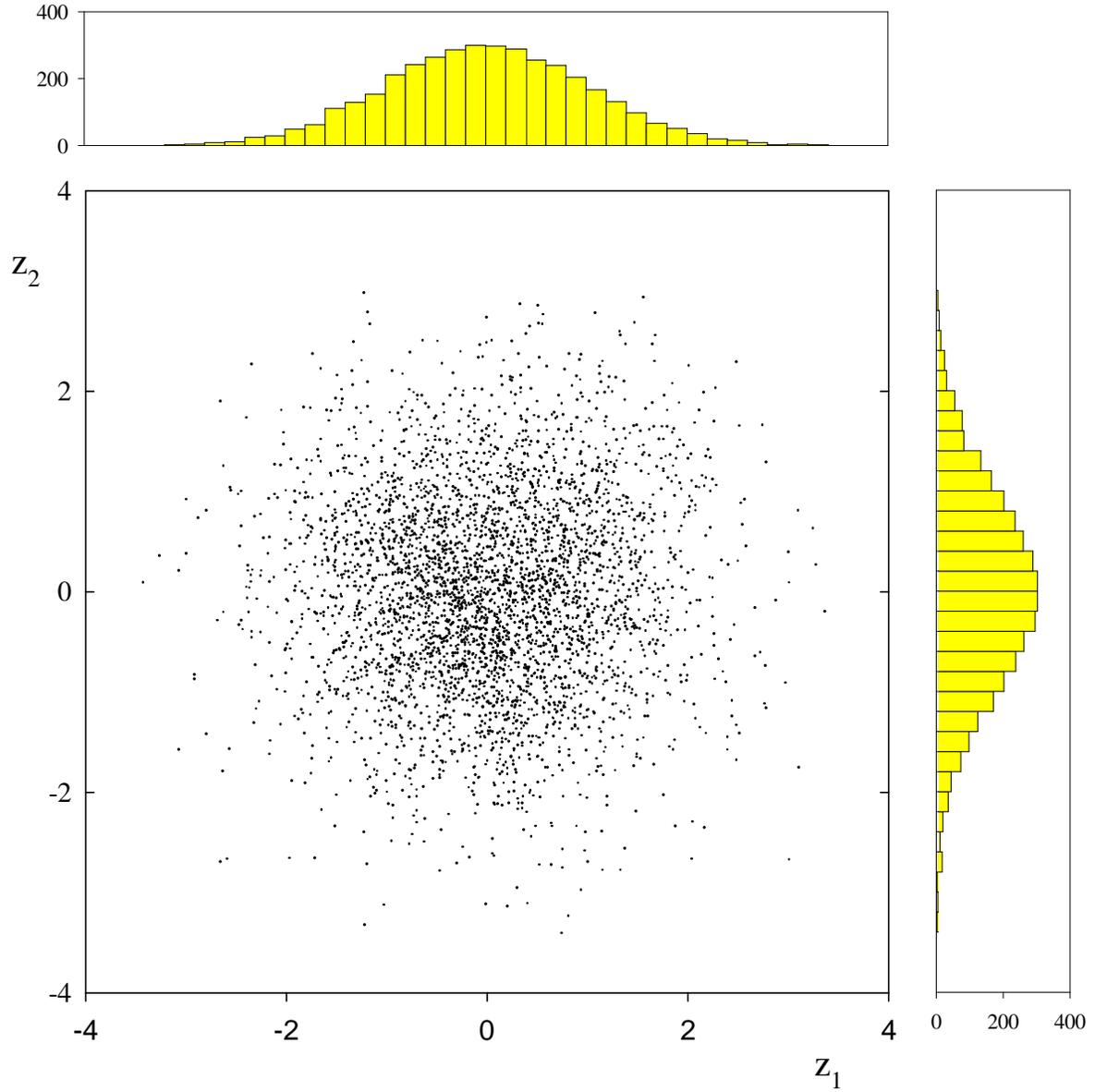}
   }
 \end{center}
\caption[foo]{\label{FIG:backtrans}
The points show the distribution of the 4000 background events after
being transformed according to Eq.\ (\ref{EQ:transform}).
The correlation coefficient is 0.03, and the two variables appear
to be uncorrelated.
}
\end{figure}

A separate transformation of the
background control sample gives the distribution shown in
Fig.\ \ref{FIG:backtrans}, which has a correlation coefficient
of 0.03.
Note that a small linear correlation coefficient does not
necessarily imply that the variables are uncorrelated.
In this case the 2-dimensional distribution is
well described by 2-dimensional Gaussian, as shown in
Fig.\ \ref{FIG:wdist}.

Since the PCA method gives a relatively good approximation of the
signal and background
probability distributions, an efficient event classification scheme
can be developed, as described in section \ref{SECT:class}.
Care needs to be taken, however, so that
the estimation of the overall efficiency and
purity of the selection is not biased.
In this example, the approximate
signal PDF is defined by 81 parameters (two projections of
40 bins, and one correlation coefficient) derived from the
4000 events in the signal control sample.
These parameters will be sensitive to the statistical fluctuations
in the control sample, and thus
if the same control sample is used to optimize the selection and
estimate the efficiency and purity, the estimates may be biased.
To reduce this bias, additional samples are generated with the
method described at the end of section \ref{SECT:pca}.
These samples are used to define the 81 parameters, and the
event classification scheme is applied to the
original control samples to estimate the purity and efficiency.
In this example data analysis, the bias is small. 
When the original control sample is
used to define the 81 parameters, the optimal signal to noise is achieved
with an efficiency of $0.880$ and purity of $0.726$.
When the PCA generated samples are used instead, 
the selection efficiency is reduced to $0.873$,
for the same purity.

When the classification scheme is
applied to the data sample, 261 events are classified as signal events.
Given the efficiency and purity quoted above,
the number of signal events in the sample is estimated to be $217 \pm 19$.

\begin{figure}
 \begin{center}
   \mbox{
    \epsfysize=12.0cm
    \epsffile[80 244 491 547]{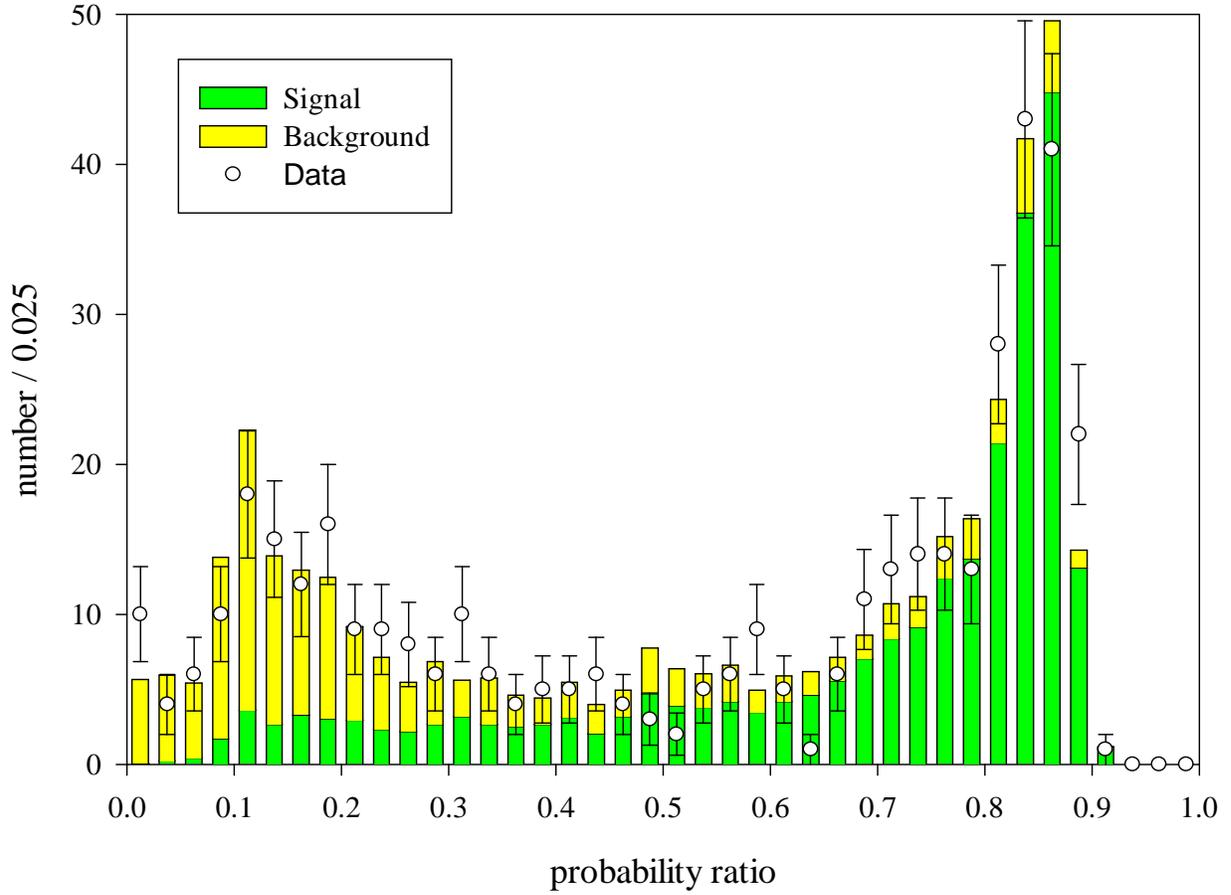}
   }
 \end{center}
\caption[foo]{\label{FIG:datatest}
A check is made
that the data sample is consistent with the model used in the
maximum likelihood analysis. The distribution of the probability ratio, 
Eq.\ (\ref{EQ:probrat}), is shown for the data events and compared to the
expected distribution, as given by a mixture of 
PCA generated signal and background samples. 
The agreement is good, the value for
$\chi^2$ is 36 for 35 degrees of freedom.
}
\end{figure}

The number of signal events in the data sample can be more accurately
determined by using a maximum likelihood analysis.
The likelihood function is defined by
\begin{equation}
L = \prod_{j=1}^{400} (f_s\, P_s({\bf x}_j) + (1-f_s)\,P_b({\bf x}_j))
\label{EQ:likelihood}
\end{equation}
where the product runs over the 400 data events,
$f_s$ is the fraction of events attributed to signal, and
$P_s$ and $P_b$ are the PCA approximated PDFs, defined by
Eq.\ (\ref{EQ:pca}).
The signal fraction, estimated by maximizing the likelihood,
is $0.617 \pm 0.040$, a relative uncertainty of 6.4\% compared to
the 8.5\% uncertainty from the counting method.
To check that the data sample is well described by the model used to
define the likelihood function, Eq.\ (\ref{EQ:likelihood}),
the ratio of probabilities, Eq.\ (\ref{EQ:probrat}),
is shown in Fig.\ \ref{FIG:datatest}, and compared to
a mixture of PCA generated signal and background samples.

\section{Fortran Implementation}

The source code for a FORTRAN-77 implementation of the methods 
described in this paper is available from the author.
The program was originally developed 
for use in an analysis of data from OPAL,
a particle physics experiment located at CERN,
and makes use of the CERNLIB library \cite{REF:cernlib}.
An alternate version is also available, in which the
calls to CERNLIB routines are replaced by calls to 
equivalent routines from NETLIB \cite{REF:netlib}.


%
%

%
%

\end{document}